\documentclass[12pt,preprint]{article}
\usepackage{epsf}
\newcommand{\eqb}{\begin{equation}}
\newcommand{\eqe}{\end{equation}}
\newcommand{\dmb}{\begin{displaymath}}
\newcommand{\dme}{\end{displaymath}}
\newcommand{\pd}{\partial}
\newcommand{\ep}{\varepsilon}
\newcommand{\eab}{\begin{eqnarray}}
\newcommand{\eae}{\end{eqnarray}}
\newcommand{\ra}{\right\rangle}
\newcommand{\la}{\left\langle}
\newcommand{\e}{\mbox{e}}
\newcommand{\be}{\begin{equation}}
\newcommand{\ee}{\end{equation}}

\newcommand{\La}{\Lambda}

\begin{document}
\begin{titlepage}
\begin{flushright}
2003-04 \\
HD-THEP-03-20\\
\end{flushright}
\vspace{0.6cm}

\begin{center}
\Large{{\bf Thermodynamic gauge-theory cascade}}

\vspace{1cm}

Ralf Hofmann

\end{center}
\vspace{0.3cm}

\begin{center}
{\em 
Institut f\"ur Theoretische Physik\\ 
Universit\"at Heidelberg\\ 
Philosophenweg 16, 69120 Heidelberg, Germany}\vspace{0.5cm}

\end{center}
\vspace{0.5cm}

\begin{abstract}

It is proposed that the cooling of a 
thermalized SU($N$) gauge theory can be formulated in terms of 
a cascade involving three effective theories with successively reduced 
(and spontaneously broken) gauge symmetries, SU($N$)$\,\to\,$U(1)$^{N-1}$\,$\to$\,Z$_N$. 
The approach is based on the 
assumption that away from a phase transition the 
bulk of the quantum interaction inherent to the system 
is implicitly encoded in the (incomplete) 
classical dynamics of a collective part made of low-energy 
condensed degrees of freedom. The properties of (some of the) statistically fluctuating fields are 
determined by these condensate(s). This leads to a quasi-particle 
description at tree-level. It appears that radiative corrections, which are 
sizable at large gauge coupling, do not change the tree-level picture 
qualitatively. The thermodynamic self-consistency of the quasi-particle approach 
implies nonperturbative 
evolution equations for the associated masses. The temperature dependence of these masses, in turn, 
determine the evolution of the 
gauge coupling(s). The hot gauge system approaches 
the behavior of an ideal gas of massless gluons at asymptotically large temperature. 
A negative equation of state is possible at a stage 
where the system is about to settle 
into the phase of the (spontaneously broken) 
Z$_N$ symmetry.

\end{abstract} 

\end{titlepage}

\section{Introduction}

The behavior of Quantum Chromodynamics (QCD) at high temperature has 
been an object of intense theoretical
study over the last decade. For a large part this effort was 
driven by the fascinating possibility to
experimentally generate a new state of matter -- the quark-gluon plasma. Owing to 
the asymptotic freedom of QCD \cite{WilczekGross} the expectations 
were high that the physics of this 
plasma would be accessible to thermal perturbation theory (TPT) for an 
experimentally realistic range of temperatures above the 
deconfinement transition. It was soon discovered that 
the perturbative expansion of thermodynamic potentials is in principle 
limited to a finite (nonanalytic) order in $\alpha_s$, see \cite{Iancu} 
and refs. therein. 
At asymptotically high temperature $T\gg T_c\sim \La_{QCD}$ 
TPT converges well in 
calculations of the thermodynamic potentials. For realistic 
temperatures, however, the convergence of the 
naive perturbative corrections to the ideal-gas behavior is poor. 
Typically, one obtains an alternating behavior which 
destroys the predictivity of TPT. Hints on the presence 
of large nonperturbative effects in the deconfined phase 
are due to lattice simulations of the thermodynamic
potentials. For example, the lattice results for the 
QCD pressure at $T\sim 4\, T_c$ are typically 20\% lower than the 
Stefan-Boltzmann limit \cite{lattice}. 
To tackle these nonperturbative effects a combination of dimensionally reduced TPT and 
lattice QCD already turned out to be fruitful \cite{Philipsen}. Hopefully it 
will generate an even deeper 
understanding of the hot gauge dynamics in the future. At the time being, however, 
the only genuine predictive but unfortunately not very insightful 
approach seems to be a pure 
lattice analysis. The present status of our 
understanding of hot (3+1)D gauge dynamics 
is thus not overly satisfactory. 

In this paper we propose a cascade-type effective theory for hot gauge 
dynamics with ingredients that are genuinely nonperturbative. 
In this sense several ideas have been put forward in the past. An approach, 
which focuses on the nonperturbative aspects of hot gauge dynamics 
around the deconfinement transition, was advocated in \cite{Pisarski}. 
It is based on the Polyakov 
loop variable as the relevant degree of freedom in a dimensionally reduced 
3D field theory. The Lagrangian 
of this theory is constrained by the demand of 
renormalizability and the global Z$_3$ symmetry. Under a number of assumptions 
an estimate of the Polyakov-loop correlator 
predicts the change of mass ratios across 
the transition. A discussion of a hot (2+1)D Georgi-Glashow model 
was carried out in \cite{KoganDunne}. 

To set up some of our conventions we now give a brief overview on the scenario 
which emerges in the present work. We consider the underlying 
dynamical principle to be a hot (3+1)D SU($N$) Yang-Mills theory. 
In this introduction we start our discussion at low temperature. 

As in \cite{KoganDunne} 
the confining phase (referred to as the center phase) 
is characterized by condensed magnetic vortices. 
A phase transition takes place at a critical 
temperature $T^c_{M\leftrightarrow C}$ where the 
magnetic gauge couplings $g_n$, belonging to an 
Abelian Higgs model valid within some range of 
temperatures above $T^c_{M\leftrightarrow C}$, diverge. Typical Nielson-Oleson 
vortices of this Abelian Higgs model, which are quasi-classical objects for temperature sufficiently larger than 
$T^c_{M\leftrightarrow C}$, become zero-energy 
defects for $g_n\to\infty$ and thus condense \cite{NO}. 
The divergence of the gauge couplings $g_n$ at $T^c_{M\leftrightarrow C}$ 
implies that gauge bosons decouple 
from the dynamics for 
$T<T^c_{M\leftrightarrow C}$. Thus the effective theory, valid for
temperatures below $T^c_{M\leftrightarrow C}$, has no continuous gauge symmetry. 
We refer to the phase described by the 
Abelian Higgs model as the magnetic phase. The validity of a magnetic description in a range 
$T^c_{M\leftrightarrow C}<T<T^c_{M\leftrightarrow M}$ derives from the fact 
that the gauge coupling $e$ in an electric phase diverges when the 
critical temperature $T^c_{E\leftrightarrow M}>T^c_{M\leftrightarrow C}$ 
is approached from above. 
The effective electric description assumes that the underlying SU($N$) theory is 
supplemented by an adjoint Higgs field $\phi$, which breaks SU($N$) maximally, 
SU($N$)$\,\to$\,U(1)$^{N-1}$. On the one hand, during the blow-up of $e$ tree-level massless 
gauge bosons of the electric phase remain tree-level massless and tree-level massive gauge bosons
decouple due to $m\propto e|\phi|$. This phenomenon renders 
the theory Abelian. On the other hand, magnetic monopoles, which are quasi-classical objects 
deep inside the electric phase, become massless and thus condense. 
For finite values of the magnetic couplings $g_n$ at $T<T^c_{E\leftrightarrow M}$ 
monopole condensates give mass to the 
magnetic gauge bosons. The dynamics of the electric ground state 
derived in Sec.\,3.1 is such that the vacuum expectation value (VEV) of the Higgs field modulus 
vanishes for $T\to\infty$. Asymptotically, we thus arrive 
at the fundamental pure gauge theory we assumed to be 
underlying the dynamics. 

The paper is organized as follows. 
In Sec.\,2 we set up and explain our approach. It 
rests on the assumption that SU($N$) 
thermodynamics can be expressed 
in terms of {\sl noninteracting} quasi-particles and the (incomplete) 
classical dynamics describing a condensed part -- 
the ground state of the system. In Secs.\,3 and 4 we show at tree-level 
how the thermodynamic self-consistency 
of this assumption determines the nonperturbative 
evolution with temperature of the gauge couplings 
in the electric and in the magnetic phase. For the electric phase 
we briefly investigate in Sec.\,3.3 how radiative corrections give 
masses to the tree-level massless gauge 
bosons in the strong-coupling limit. A discussion of the confining phase, where the discrete subgroup 
Z$_N$ is broken spontaneously, is carried out in Section 5. 
In Section 6 we match the three different 
phases using Clausius-Clapeyron 
conditions. As a result our approach implies that SU($N$) thermodynamics 
is parameterized by a single mass scale. A summary and an outlook on future work 
is given in Sec.\,7. In addition, some aspects of 
cosmological phase transitions are discussed in the light of our approach.

\section{General aspects of the construction}

In this section we explain our approach 
to hot SU($N$) gauge dynamics. We start with the fundamental assumption 
that away from a phase transition quantum interactions can be expressed in 
terms of {\sl noninteracting} quasi-particle statistical fluctuations, that is, free-particle 
modes with a $T$ dependent mass, and in terms of the classical Euclidean 
dynamics of charged scalar Higgs field(s) $\{\phi\}$ describing a 
condensed part of the system. The latter 
represents the ground state of the theory at given $T$. In the following we consider the situation 
that the effective theory 
describing SU($N$) thermodynamics in some temperature range 
has a continuous gauge symmetry\footnote{In Sec.\,5 the case of a discrete symmetry will be discussed.}. 
In such an effective description gauge fields $\{A_\mu\}$ can be present in an explicit form. 
They also can appear implicitly as 
condensed degrees of freedom (CDOF) which are 
defined in terms of $\{A_\mu\}$. At a given temperature the CDOFs build the 
ground state whose dynamics is 
described classically. We will see later 
that the assumption of a nonfluctuating, classical part in the system 
turns out to be self-consistent. Let us discuss this part 
in more detail. First we consider a situation where explicit gauge fields 
are absent. In this hypothetic case the scalar fields $\{\phi\}$ represent CDOFs without residual 
gauge field interactions. Condensation may occur in the limit where 
the to-be-condensed degrees of freedom carry no energy. Consequently, 
the condensate would have vanishing energy density. In a 
macroscopic approach, where the resolution is comparable with temperature, 
this is a condition which must be imposed. We will refer to it as 
zero-energy condition (ZEC). Scalar field 
configurations, which solely depend on the Euclidean time coordinate $x_4$, carry no 
energy density if they are Bogomol'nyi-Prasad-Sommefield (BPS) \cite{BPSaut} saturated solutions to 
the scalar-sector equations of motion. To satisfy the ZEC nontrivially in thermal 
equilibrium we thus have to construct a potential $V$ 
for the fields $\{\phi\}$ which permits periodic and BPS saturated solutions. 
In addition, the solutions $\{\phi\}$ must yield an $x_4$ independence of $V$. 
The required potential is of the $\phi^{-2}$-type. 

Next we let explicit gauge fields enter the stage. In a thermalized system no spatial 
direction is singled out. The part of the system being represented by 
statistically fluctuating (quasi-)particles already satisfies this requirement. In consequence, 
the {\sl ground state} ought not carry any gauge field curvature. This situation 
will be referred to as zero-curvature condition (ZCC) in the following. A gauge field 
configuration, which is pure gauge and a solution to the gauge field 
equations of motion in the background of the 
scalar-field configuration, is needed. According to the gauge field 
equations of motion it is necessary for the existence of these pure gauge configurations 
that source terms vanish. This happens if 
gauge covariant derivatives annihilate the scalar backgrounds.
As a result, the ZEC for the ground state, which was imposed 
in the absence of explicit gauge fields, is violated. This 
phenomenon has the following interpretation: The energy density of the ground state, 
given by the potential $V(\{\phi\})$, 
is generated by the CDOF interactions which are mediated by 
explicit gauge fields. The alert reader may object that our ground state configurations 
are no longer exact solutions to the Euclidean 
equations of motion since the gauge field equations 
were solved in the {\sl background} of the 
configurations $\{\phi\}$ which, in turn, were obtained by solving 
scalar-sector BPS equations at $\{A_\mu\}=0$. The influence of the explicit gauge 
fields on the scalar field configurations is thus ignored. In fact, viewing the ground state as 
a thermalized system by itself leads to a contradiction 
to the fundamental thermodynamic relation for total pressure $p$ and the total 
energy density $\ep$
\eqb
\label{demo}
\ep(T)=T\frac{dp(T)}{dT}-p(T)\,.
\eqe
Relation (\ref{demo}) would directly follow from the (hypothetic) 
partition function for such a system \cite{Gorenstein}. We conclude 
that configurations, which satisfy the requirements of ZEC and ZCC but, at the same time, 
do not solve the Euclidean equations of motion completely, are thermodynamically 
inconsistent. This is the point 
where the fluctuating part of the system comes in. It will be demonstrated in Sections 3.2, 4.2, and 5.2 
that the above-mentioned $\phi^{-2}$-type potentials for the Higgs-fields 
imply large masses ($\gg T$) for the collective modes of the CDOFs. 
These modes are thus of no 
statistical relevance. In consequence, we are left with gauge-field QPEs. 
At tree-level their masses $\{m\}$ are proportional to the product of 
a gauge-coupling and a Higgs-field 
modulus which both can be small. 

Let us now come back 
to the problem with the incomplete ground-state `solution' to 
the classical equations of motion. It is physically tempting to 
interprete the reaction of the condensed 
sector, represented by the Higgs-field configurations, on the presence of 
zero-curvature gauge-field configurations in terms of 
emission and absorption processes of gauge-field 
quasi-particle excitations (QPEs). This renders the condensed 
part a heat reservoir for the fluctuating gauge fields! In each 
emission process the ground state inherits some of its 
properties to the QPE in the form of a $T$ dependent mass $m$. 
This picture is thermodynamically self-consistent (TSC) if 
the conditions \cite{Gorenstein} 
\eqb
\label{TSCgen}
\pd_{\{m\}}p=0\,
\eqe
are satisfied. Notice that in (\ref{TSCgen}) the pressure $p$ 
has a ground-state component and a component due to QPEs. 
A tree-level consequence of (\ref{TSCgen}) 
is the prediction of the nonperturbative 
evolution of the gauge couplings\footnote{Radiative corrections do not change the qualitative 
picture, see Sec.\,3.3.} in the cooling system. When some critical temperature $T^c$ is approached 
the couplings diverge at some finite Higgs-field modulus. This means that the tree-level 
massive gauge bosons, whose mass is proportional to the gauge coupling times the Higgs-VEV 
modulus decouple from the dynamics. It also implies 
that topological defects, whose action scales with an 
inverse power of the gauge coupling, start 
to behave extremely quantum mechanically. They condense and form the ground state 
of the subsequent phase of reduced (gauge) symmetry at lower temperature.   
\begin{figure}
\begin{center}
\leavevmode
\leavevmode
\vspace{5cm}
\includegraphics{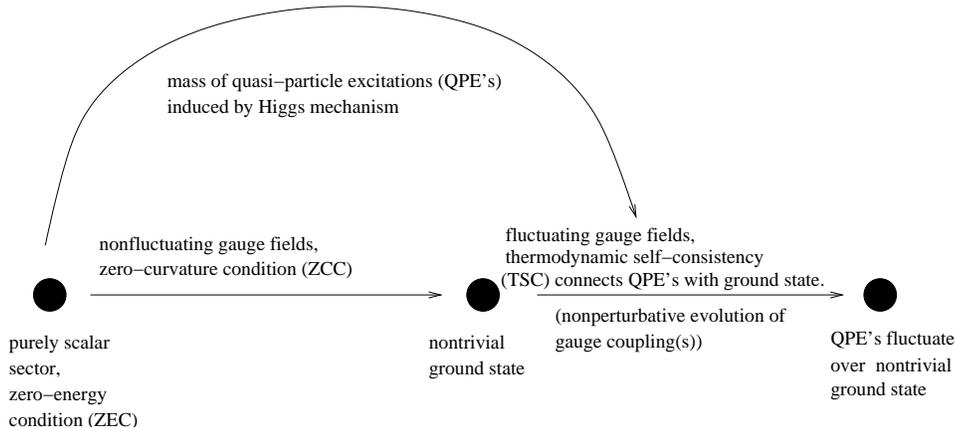}
\end{center}
\caption{The construction of a 
thermodynamically self-consistent, effective 
gauge theory with a nontrivial ground-state structure 
and quasi-particle excitations. 
\label{Fig0}}   
\end{figure}
We have summarized our construction graphically in Fig.\,1. 

How does thermal perturbation theory comply with this picture? 
At high temperature one naively expects it to yield a good description 
due to asymptotic freedom. It is well-known, however, 
that a straightforward expansion in powers 
of $\alpha_s$ can be carried out in perturbation theory only up to a finite order 
due to the magnetic screening problem. 
Moreover, a perturbative expansion of the thermodynamic potentials converges 
only poorly even at temperatures considerable larger than $T^c$. Reorganizations of perturbation theory 
have been suggested which amount to mathematical manipulations 
of the weak coupling expansion and thus are of a mere academic value. More promising are 
variational approaches such as 
screened perturbation theory \cite{sPT}. 
Specifically, hard thermal loop perturbations theory (HTLPT) 
uses a variational mass parameter $m_D$ identified with the 
Debye screening mass in a weak coupling 
expansion in gluodynamics. For $T\ge 10\,T^c$ the approach seems 
to yield a stable result for the thermodynamic pressure 
as one goes from leading to next-to-leading 
order in the loop expansion \cite{Andersen}. 
With decreasing $T$ stability gets worse and the disagreement 
with the lattice becomes large. A matching of 
our approach with HTLPT can be performed by demanding that the typical Higgs 
induced gauge boson mass in the electric 
phase, see Sec.\,3.2, and its first derivative w.r.t. $T$ coincide 
with $m_D$ and $\frac{d m_D}{dT}$, respectively. These are two conditions 
for two unknown quanitites, namely $T_{\tiny\mbox{match}}$ 
and $\Lambda_E$, see Sec.\,3.1.      

We conclude this section with the remark that 
thermodynamic models assuming a phenomenological ground-state 
energy, a phenomenological ground-state pressure and gluonic QPEs have been 
fitted to lattice data for the total energy density, the 
total pressure, and the entropy of hot SU(2) and SU(3) 
Yang-Mills theories \cite{Gorenstein,Peshier} and hot QCD \cite{Heinz}. 
It was found in all cases that the 
common mass $m_g$ of the quasi-particles rises rapidly 
when the critical temperature of the deconfinement 
transition is approached from above -- in qualitative agreement 
with our present discussion.

\section{Very high temperature: Electric phase}

\subsection{Model and ground-state structure}

At very high temperature we consider the following Euclidean action
\eqb
\label{eucactn}
S_E=-\int_0^{x_4} d\tau\int d^3x\,\left\{\frac{1}{2}\mbox{tr}\left[
\widehat{G_{\mu\nu}G_{\mu\nu}}\right]+
\mbox{tr}\left[{\cal D}_\mu\phi{\cal D}_\mu\phi\right]+
{\cal V}(\phi)\right\}\,,
\eqe
where field strength and covariant derivative are respectively defined as
\eab
\label{covdern}
G_{\mu\nu}&=&\pd_\mu A_\nu-\pd_\nu A_\mu+ie[A_\mu, A_\nu]\,,\ 
\ \ {\cal D}_\mu\phi=\pd_\mu\phi+ie[A_\mu,\phi]\,,\nonumber\\ 
A_\nu&=&A^a_\nu t^a\,,\ \ \ \ \ (a=1,\cdots,N^2-1)\,.
\eae
The hat symbol on top of the gauge kinetic term in (\ref{eucactn}) 
indicates that certain classes of one-particle irreducible 
Feynman diagrams, which would follow from the 
usual gauge Lagrangian, are forbidden, see Sec.\,3.3. 
The hermitian and traceless matrices $t_a$ are the generators of SU($N$) in the fundamental representation. 
They are normalized as 
$\mbox{tr}\,t^a t^b=\frac{1}{2}\delta^{ab}$. We assume that SU($N$) is spontaneously broken as
\eqb
\label{breakdown}
\mbox{SU}(N)\to \mbox{U}(1)^{{N-1}}\,
\eqe
by the VEV of an adjoint Higgs field, 
\eqb
\label{adHiggs}
\phi=\phi^a t^a\,,\ \ \ \ \ \ \ \ (\phi^a\ \ \mbox{real})\,.
\eqe
In pure SU($N$) gauge 
theory such a field does not exist on the fundamental level. It is assumed here that the Higgs 
VEV of Eq.\,(\ref{adHiggs}) forms due to the condensation of topological defects. The so-called calorons 
\cite{KraalvanBaal} are possible
candidates. Calorons are nontrivial-holonomy and BPS saturated solutions 
to the classical, Euclidean Yang-Mills equations at finite
$T$. An isolated caloron thus carries no energy, and consequently 
it qualifies for a CDOF. For SU($N$) there are $N$ BPS magnetic monopole 
constituents in each caloron. They manifest themselves the stronger as lumps in the action density 
the lower $T$. The overall magnetic 
charge of a caloron is zero. The presence of constituent 
BPS monopoles is a crucial fact which we will come back to in Section 3.2. 
It is conceivable that the assumed 
condensation of the adjoint 
Higgs field (\ref{adHiggs}) is 
microscopically driven by interacting calorons and anticalorons. 
For $N>2$ a possible {\sl local} definition of $\phi^a$ 
in terms of fundamental fields is
\eqb
\label{locdefphi}
\phi^a\propto d^{abc} \la G_{\mu\nu,b}G_{\mu\nu,c}\ra\,,
\eqe
where $d^{abc}$ denotes the tensor defined by $\mbox{tr}\, t^a\{t^b,t^b\}=D d^{abc}$. 
It is also possible that the Higgs 
VEV of Eq.\,(\ref{adHiggs}) has a {\sl nonlocal} dependence on the 
fundamental fields $A_\mu$.  

A gauge bosons, which is defined w.r.t. the generator $t^a$, acquires the 
following mass-squared
\eqb
\label{massmatrix}
(m^{a})^2=-2e^2\,\mbox{tr}\,[\phi,t^a][\phi,t^a]\,.
\eqe
We will refer to the ($N-1$) gauge fields, 
which remain massless at tree-level, as tree-level massless (TLM). The modes of the 
$N(N-1)$ massive gauge fields are QPEs at tree-level. We will refer to them 
as tree-level heavy (TLH). In principle, there are in addition 
QPEs from the non-Goldstone directions of the Higgs-field 
excitations. We will show in Sec.\,3.2 that these possible excitations 
are statistically irrelevant over the entire relevant temperature range.     

Since the 
Higgs VEV $\phi$ is a hermitian (traceless) matrix we can always 
reach a gauge where $\phi$ is diagonal. We refer to this gauge as 
the diagonal gauge (DG). 
For a maximal symmetry breaking, SU($N$)$\to$ U(1)$^{N-1}$, all eigenvalues of $\phi$ must be different. 
In DG and for even $N$ a possible 
traceless Higgs VEV is
\eqb
\label{evenbreak} 
\phi=\mbox{diag}(\tilde{\phi}_1,\tilde{\phi}_2,\cdots,\tilde{\phi}_{N/2})\,,
\eqe
where 
\eqb
\label{comps2}
\tilde{\phi}_i\equiv|\tilde{\phi}_i|\frac{\tau_3}{2}\,,\ \ \ \ \ |\tilde{\phi}_i|\equiv\sqrt{2\,\mbox{tr}\,
\tilde{\phi}^2_i}\,, 
\eqe
$\tau_3$ refers to the third Pauli matrix, and 
$|\tilde{\phi}_i|\not=|\tilde{\phi}_j|>0\,,\ (i\not=j),$ are real 
numbers. For odd $N$ the Higgs VEV in DG can be taken as 
\eqb
\label{oddbreak} 
\phi=\mbox{diag}(\tilde{\phi}_1,\tilde{\phi}_2,\cdots,\tilde{\phi}_{(N-1)/2},0)\,,
\eqe
where the last entry is not a matrix but a number. 
We now relax the DG by allowing for SU(2) subgroup rotations to act 
on the matrices $\tilde{\phi}_i$. We refer to this incompletely fixed gauge as RDG. 
To avoid repetition we will address explicitly only the case when $N$ 
is even. A gauge invariant potential ${\cal V}$ 
is defined as\footnote{For odd $N$ the 
summation index and the label run up to $(N-1)/2$ in (\ref{gipotBPS}) and  
in (\ref{Vn1/2}), (\ref{persoln}),  (\ref{simbreak}), (\ref{zerocurn}), 
(\ref{configz}), respectively. 
In Eqs.\,(\ref{Vn1/2}), (\ref{persoln}), and (\ref{zerocurn}) the number
zero is the last entry on the right-hand sides. The expression to the very right 
in Eq.\,(\ref{gipotBPS}) translates into $\frac{1}{16}\Lambda_E^6 \mbox{tr} \left(\phi_b^2\right)^{-1}$ 
for odd $N$. The block matrix $\phi_b$ is defined as $\phi_b\equiv\mbox{diag}
(\tilde{\phi}_1,\tilde{\phi}_2,\cdots,\tilde{\phi}_{(N-1)/2})$. Notice that 
$\mbox{tr} \left(\phi_b^2\right)^{-1}$ is a gauge invariant quantity.} 
\eqb
\label{gipotBPS}
{\cal V}(\phi)\equiv\mbox{tr}\left[{\cal V}^
\dagger_{1/2}
{\cal V}_{1/2}\right]=\frac{1}{2}\Lambda_E^6 \sum_{i=1}^{N/2} \frac{1}{|\tilde{\phi}_i|^2}=
\frac{1}{16}\La_E^6 \mbox{tr} \left(\phi^2\right)^{-1}\,.
\eqe
In RDG we define
\eqb
\label{Vn1/2}
{\cal V}_{1/2}\equiv\Lambda_E^3\,\mbox{diag}
\left(\e^{-i\delta_1}\tau_3 \frac{\tilde{\phi}_1}{|\tilde{\phi}_1|^2},\cdots,
\e^{-i\delta_{N/2}}\tau_3 \frac{\tilde{\phi}_{N/2}}{|\tilde{\phi}_{N/2}|^2}\right)\,.
\eqe
In (\ref{gipotBPS}) and (\ref{Vn1/2}) $\La_E$ denotes a mass scale. 
According to our general approach (see Sec.\,2) we first 
disregard the gauge field sector. To describe 
the ground state in thermal equilibrium $|\tilde{\phi}_i|$ must not 
depend on any coordinate $x_\mu$ since it determines the potential in 
(\ref{gipotBPS}). Moreover, 
$\phi$ must be periodic in $x_4$. 
For a zero-energy configuration the kinetic term in the Euclidean action 
(\ref{eucactn}) is equal to the potential term. Solutions to 
the `BPS' equation 
\eqb
\label{BPSn}
\pd_{x_4}\phi={\cal V}_{1/2}
\eqe
are guaranteed to satisfy this last condition. For $\delta_i=\pm \frac{\pi}{2}$ 
(\ref{BPSn}) has periodic solutions, namely
\eab
\label{persoln}
&& \phi_{k_1,\cdots,k_{N/2}}(x_4)=\sqrt{\frac{\Lambda_E^3}{2\pi T}}\times \nonumber\\ 
&& \mbox{diag}\left(|k_1|^{-1/2}\tau_1\exp\left[2\pi ik_1 T\tau_3 x_4\right],\cdots,
|k_{N/2}|^{-1/2}\tau_1\exp\left[2\pi ik_{N/2} T\tau_3 x_4\right]\right)\,. 
\nonumber\\ 
\eae
Notice that $|\phi_{k_1,\cdots,k_{N/2}}|$ vanishes for asymptotically large $T$, 
$T\gg\La_E$. We thus recover the 
pure gauge theory. Due to asymptotic freedom perturbation theory should be applicable 
in this limit. The $T$ dependence of the solutions (\ref{persoln}) 
ensures that thermodynamic potentials asymptotically approach their 
respective ideal-gas limits. In (\ref{persoln}) the numbers $k_i$ 
denote positive or negative integers. 
Their modulus counts the number of times each SU(2) block winds around the 
pole at $\phi=0$ in (\ref{Vn1/2}) as $x_4$ runs from 
zero to $\beta$. Each block can be gauge 
rotated to the form in (\ref{evenbreak}) by SU(2) subgroup transformations. 
Consequently, for the solution (\ref{persoln}) to 
break SU($N)\to\mbox{U}(1)^{{N-1}}$ 
we have to impose 
\eqb
\label{break}
|k_i|\not=|k_j|\,,\ (i\not=j)\,.
\eqe
The minimal way to satisfy (\ref{break}) in view of the resulting potential is
\eqb
\label{simbreak}
k_1=1\,,\ k_2=2\,,\ \cdots\,,\ k_{N/2}=N/2\,.
\eqe
Eqs.\, (\ref{simbreak}) imply 
\eab
\label{potonsol}
{\cal V}(|\phi_{1,\cdots,{N/2}}|)&\equiv&B_E(T)=\frac{\pi}{8}\,N(\frac{N}{2}+1)\,\Lambda_E^3 T
\,\ \ \ \
 (\mbox{even}\ N)\,,\nonumber\\ 
{\cal V}(|\phi_{1,\cdots,{(N-1)/2}}|)&\equiv&B_E(T)=\frac{\pi}{16}\,(N^2-1)\,\Lambda_E^3 T
\,\ \ \ \ (\mbox{odd}\ N)\,.
\eae
Next explicit gauge fields may enter the stage. The 
ground state of the system is described 
by a zero-curvature configuration, $G_{\mu\nu}\equiv 0$. 
Taking $\phi_{1,\cdots,{N/2}}$ as a background in the 
equations of motion for the gauge fields,
\eqb
\label{eomgaug}
{\cal D}_\mu G_{\mu\nu}=2ie[{\cal D}_\nu\phi,\phi]\,,
\eqe
is consistent with zero curvature: For $G_{\mu\nu}$ to be zero it is necessary that 
the right-hand side of (\ref{eomgaug}) vanishes. This happens 
if ${\cal D}_\nu\phi=0$. For the background $\phi_{1,\cdots,{N/2}}$ it is easily checked 
that the pure-gauge configuration 
\eqb
\label{zerocurn}
A^{1,\cdots,{N/2}}_{\nu}=\delta_{\nu 4}\frac{4\pi T}{e}\,\mbox{diag}\,\left(\frac{\tau_3}{2},
2\,\frac{\tau_3}{2},\cdots,\frac{N}{2}\,\frac{\tau_3}{2}\right)\,
\eqe
satisfies the condition ${\cal D}_\nu\phi_{1,\cdots,{N/2}}=0$. Therefore $A^{1,\cdots,{N/2}}_{\nu}$ 
solves the equations 
of motion (\ref{eomgaug}) at zero curvature. When 
evaluated on the configurations 
\eqb
\label{configz}
\left(\phi_{1,\cdots,{N/2}}; A^{1,\cdots,{N/2}}_{\nu}\right)\,.
\eqe
the action density in (\ref{eucactn})   
reduces to the 
potential (\ref{potonsol}). 

\subsection{Thermal self-consistency and quasi-particles}

To decide which QPEs are statistically 
relevant the masses of the associated fields have to be estimated. 
In this section we will be content 
with a tree-level analysis. QPEs fluctuate about the configuration (\ref{configz}). 
On the one hand, we find at even $N$ for non-Goldstone like scalars
\eqb
\label{smassE}
\frac{m_{s,k}}{T}\equiv\sqrt{\pd^2_{|\tilde{\phi}_k|}{\cal V}}/T=\sqrt{12}\pi\,k\,,\ 
\ \ \ (k=1,\cdots, \frac{N}{2})\,.
\eqe
Eq.\,(\ref{smass}) indicates that non-Goldstone like fluctuations 
are strongly Boltzmann suppressed. They can safely be neglected in the following. 
On the other hand, for TLH gauge bosons the mass-to-temperature ratios 
\eqb
\label{vmass}
\frac{m^{TLH}_{E,\alpha}}{T}\,,\ \ \ \ (\alpha=1,\cdots, N(N-1))
\eqe
depend linearly on the 
gauge coupling modulus $|e|$ and on the ratios $\frac{|\tilde{\phi}_k|}{T}$. 
Both quantities can be small, and thus gauge modes are possibly fluctuating. In DG, where 
$\phi_{rs}=\phi_r\delta_{rs}\,,\ (r,s=1,\cdots,N)$, 
we may define the $N(N-1)$ normalized generators, which correspond to the TLH gluons, as follows
\eab
\label{massgen}
t^{IJ}_{rs}&=&\frac{1}{2}\left(\delta_r^I\delta_s^J+\delta_s^I\delta_r^J\right)\,,\ \ \ \ \ \ \ 
\tilde{t}^{IJ}_{rs}=-\frac{i}{2}\left(\delta_r^I\delta_s^J-\delta_s^I\delta_r^J
\right)\,,\nonumber\\ 
& & \ \ \ \ \ \ \ \ \ \ \ \ \ \ \ \ \ \ \ \ \ \ \ \ \ \ \ \ \ \ \ \ \ (I=1,\cdots,N,\ J>I)\,.
\eae
Substituting (\ref{massgen}) into (\ref{massmatrix}) yields 
\eqb
\label{gernalmasss}
(m^{IJ})^2=e^2\,(\phi_I-\phi_J)^2\,.
\eqe
for both generators $t^{IJ}$ and $\tilde{t}^{IJ}$. According to (\ref{gernalmasss}) this leads 
to two-fold and four-fold degeneracies of 
gauge boson masses in the spectrum. The lowest mass $m^{TLH}_E$ is 
\eqb
\label{vmasssimE}
m^{TLH}_E=|e|\left\{\begin{array}{c} \sqrt{\frac{\La_E^3}{\pi T N}}\left(\frac{1}
{\sqrt{1-\frac{2}{N}}}-1\right)\,,
\ \ \ \ \ \ \ \ \ \ \ (\mbox{even}\ N)\nonumber\\ 
\sqrt{\frac{\La_E^3}{\pi T (N-1)}}\left(\frac{1}
{\sqrt{1-\frac{2}{N-1}}}-1\right)\,,\ \ \ \ (\mbox{odd}\ N)\,\end{array}\right.\,.
\eqe
For simplicity we will set all masses $m^{TLH}_{E,\alpha}$ equal to $m^{TLH}_E$ 
in the following. Using (\ref{gernalmasss}) it is straightforward to take TLH 
masses into account exactly for $N$ given. When estimating the pressure component 
arising from TLH modes the approximation $m^{TLH}_{E,\alpha}=m^{TLH}_E$ yields an 
upper bound. 
       
\noindent The total pressure of the system is
\eqb
\label{press}
p_E\equiv T\frac{d \ln Z_E}{dV}\,
\eqe
where $Z_E$ denotes a partition function, 
\eqb
\label{ZE}
Z_E\propto\mbox{Tr}_{\tiny{\mbox{periodic}}} \exp\left[S_E^{\tiny{\mbox{free}}}-\frac{B_E V}{T}\right]\,,
\eqe
and $V$ is the (infinite) three-dimensional 
(box-)volume of the system. The action $S_E^{\tiny{\mbox{free}}}$ decomposes into a part for the $N(N-1)$ 
noninteracting TLH gauge bosons and a part for the $(N-1)$ noninteracting TLM gauge bosons. 
The energy density of the ground-state, $B_E(T)$, is defined in (\ref{potonsol}). 
The total pressure $p_E$ follows from (\ref{press}) and (\ref{ZE}) as 
\eqb
\label{pressEE}
p_E=p^{id}_E-B_E(T)\,.
\eqe
The total energy density $\ep_E$
\eqb
\label{quae}
\ep_E=T\frac{dp_E(T)}{dT}-p_E(T)\,
\eqe
assumes the same form as in the case of $B_E=\,\,$const, $m^{TLH}_E=\,\,$const, 
\eqb
\label{idgasB}
\ep_E\equiv\ep^{id}_E+B_E(T)\,,
\eqe
provided that the condition 
\eqb
\label{sscE}
\frac{\pd p_E}{\pd m^{TLH}_E}=\frac{\pd p^{id}}{\pd m^{TLH}_E}-\frac{\pd B_E}{\pd m^{TLH}_E}=0
\eqe
is satisfied \cite{Gorenstein}. In (\ref{pressEE}) and (\ref{idgasB}) 
$p^{id}_E(\ep^{id}_E)$ denote the pressure (energy density) 
of an ideal gas of $(N-1)$ TLM gluons (2 polarizations) and $N(N-1)$ TLH gluons (3 polarizations)
\eab
\label{pid,eid}
p^{id}_E&\equiv&-\frac{T^4}{2\pi^2}\left[2(N-1)P(0)+3N(N-1)P(a^{TLH}_E)\right]\,,\nonumber\\ 
\ep^{id}_E&\equiv&\frac{T^4}{2\pi^2}\left[2(N-1)E(0)+3N(N-1)E(a^{TLH}_E)\right]\,,\ \ \ 
a^{TLH}_E\equiv\frac{m^{TLH}_E}{T}\,,\nonumber\\ 
P(a)&\equiv&\int_0^\infty dx\,x^2\log\left[1-\exp(-\sqrt{x^2+a^2})\right]\,,\nonumber\\ 
E(a)&\equiv&\int_0^\infty dx\,x^2\frac{\sqrt{x^2+a^2}}{\exp(\sqrt{x^2+a^2})-1}\,.
\eae
Eq.\,(\ref{sscE}) expresses the TSC of the approach. 
Substituting  (\ref{potonsol}) and (\ref{pid,eid}) in (\ref{sscE}) 
and performing a few elementary manipulations yields 
\eqb
\label{ladiffE}
\frac{d\lambda_E}{da^{TLH}_E}=-\frac{12}{(2\pi)^6}\lambda_E^4\,a^{TLH}_E\,D(a^{TLH}_E)\left\{\begin{array}{c} 
\frac{N-1}{N+2}\,,
\ \ \ \ (\mbox{even}\ N)\nonumber\\ 
\frac{N}{N+1}\,,\ \ \ \ (\mbox{odd}\ N)\,\end{array}\right.\,.
\eqe
The following definitions have been made:
\eqb
\label{defi}
T_E\equiv\frac{\Lambda_E}{2\pi}\,,\ \ \ \lambda_E\equiv \frac{T}{T_E}\,, \ \ \ 
D(a)\equiv\int_0^{\infty} dx\,\frac{x^2}{\sqrt{x^2+a^2}}\,\frac{1}{\exp\left[\sqrt{x^2+a^2}\right]-1}\,.
\nonumber\\ 
\eqe
The evolution equation (\ref{ladiffE}) can be solved numerically. 
Fig.\,2 shows a plot of the function $a\,D(a)$ which appears on the 
right-hand side of Eq.\,(\ref{ladiffE}). This function is peaked at 
$a=1$, and it is positive definite. As a consequence, the right-hand side of (\ref{ladiffE}) is negative definite. 
It is thus possible to invert $\lambda_E=\lambda_E(a^{TLH}_E)$ into a function $a^{TLH}_E(\lambda_E)$.
Notice that in deriving (\ref{ladiffE}) we have assumed that the 
gauge coupling $e$ depends on $T$ only {\sl implicitly} 
via $a^{TLH}_E$.
\begin{figure}
\begin{center}
\leavevmode
\leavevmode
\vspace{5cm}
\includegraphics{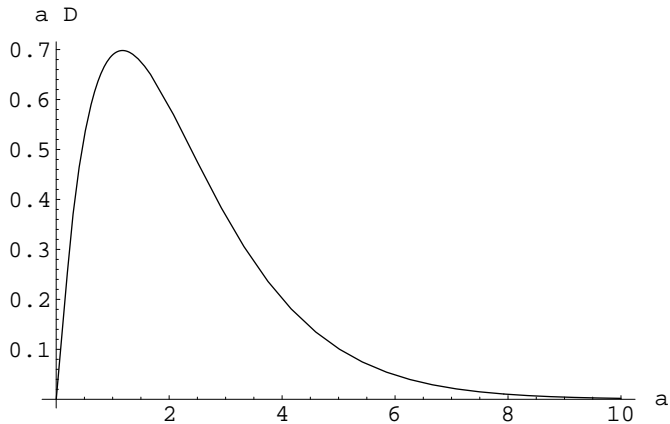}
\end{center}
\caption{The function $a\,D(a)$
\label{Fig1}.}   
\end{figure}
The definitions for $a^{TLH}_E$ in (\ref{pid,eid}) and for $\lambda_E$ in (\ref{defi}) 
can be inserted into (\ref{vmasssimE}), and the result can be solved for the 
gauge-coupling modulus $|e|$. We arrive at the following $\lambda_E$ dependence
\eqb
\label{Tdepe}
|e|(\lambda_E)=\sqrt{\frac{\lambda^3_E}{8 \pi^2}}\,\,a^{TLH}_E(\lambda_E)\,\left\{\begin{array}{c} 
\sqrt{N}\left(\frac{1}
{\sqrt{1-\frac{2}{N}}}-1\right)\,,
\ \ \ \ \ \ \ \ \ \ \ (\mbox{even}\ N)\nonumber\\ 
\sqrt{N-1}\left(\frac{1}
{\sqrt{1-\frac{2}{N-1}}}-1\right)\,,\ \ \ \ (\mbox{odd}\ N)\,\end{array}\right.\,.
\eqe
For $N=10$ we will discuss the solutions to (\ref{ladiffE}) more carefully in Sec.\,6. 
At this point it is only important 
to notice that for a given value of $N$ and given 
initial conditions $(\lambda^i_E,a^i_E)$ the inverted solution 
$a^{TLH}_E(\lambda_E)$ diverges at some finite, critical value $\lambda^c_E<\lambda^i_E$. 
We associate this value with 
the critical temperature 
\eqb
\label{Tcem}
T^c_{E\leftrightarrow M}=\frac{\La_E}{2\pi}\lambda^c_E\,.
\eqe
According to (\ref{Tdepe}) the behavior 
\eqb
\label{dive}
a^{TLH}_E\to\infty \ \ \ \ \mbox{as}\ \ \ \ 
T\searrow T^c_{E\leftrightarrow M}
\eqe
implies that also the gauge coupling $e$ 
diverges for $T\searrow T^c_{E\leftrightarrow M}$! 

Our approach to the ground-state dynamics was 
a macroscopic, thermodynamic one. On a microscopic level, that is, when probing the 
system with momenta larger than $T$, the ground state no longer is spatially homogeneous. We 
already pointed out that a good candidate for the CDOFs are calorons. Calorons 
have BPS monopole constituents. The mass of a singly charged BPS magnetic monopole 
of an SU(2) embedding labeled by $k$ is given as \cite{PrSom}
\eqb
\label{massmon}
M_{mon}=\frac{4\pi}{|e|}|\tilde{\phi}_k|\,.
\eqe
In (\ref{massmon}) $|\tilde{\phi}_k|$ denotes the modulus of one of the SU(2) blocks 
which were obtained in (\ref{persoln}). Obviously, for 
$T\searrow T^c_{E\leftrightarrow M}$ the mass of the magnetic monopole vanishes, 
$M_{mon}\to 0$. This renders monopoles, which are released by the calorons at $T^c_{E\leftrightarrow M}$, 
good candidates for CDOFs. We expect a transition from the monopole uncondensed to the monopole condensed phase 
to occur at $T^c_{E\leftrightarrow M}$.       

\subsection{Radiatively induced mass for TLM modes} 

We have relied on a tree-level 
analysis in Sec.\,3.2. According to our fundamental 
assumption all effects of the interaction between QPEs are encoded in 
their $T$ dependent masses. This constraint ought to be 
maintained on the quantum level. Starting from a one-particle 
{\sl reducible} self-energy diagram, the generation of a 1PI self-energy diagram by additional gauge boson 
exchange(s) is forbidden. This 1PI diagram would 
double count the interaction in our 
approach based on QPEs. One may formulate the constraint in a more 
eidetic way: The exchange of a gauge boson between 
two clusters in a one-particle 
{\sl reducible} diagram would correspond to a consideration of 
long-range correlations. These correlations are already accounted for by the ground-state dynamics which 
is responsible for the TLH masses at tree-level.   

We have used in Sec.\,3.2 and will exemplarily demonstrate in Sec.\,7 
that the gauge coupling $e$ is weak for $T\gg T^c_{E\leftrightarrow M}$ and that 
there is a strong coupling regime for $T\sim T^c_{E\leftrightarrow M}$. 
At weak coupling radiative corrections to the tree-level masses 
are of the usual, (resummed) perturbative screening type. 
At lowest order in $|e|$ we have 
\eqb
\label{scmass}
\Delta m_{sc}\sim \sqrt{N}|e|\,T\,.
\eqe
For $|e|\ll 1$ this correction is 
thermodynamically irrelevant. At large coupling, $|e|\gg 1$, the masses of 
TLH modes are large. Thus TLH modes are 
thermodynamically irrelevant. TLH quantum fluctuations 
do radiatively induce masses for the TLM modes. 
Recall that TLM gauge bosons live in the Cartan 
sub-algebra. Thus they do not have a 
vertex with one another. They can only couple by means of intermediate 
TLH particles. 
\begin{figure}
\begin{center}
\leavevmode
\leavevmode
\vspace{5cm}
\includegraphics{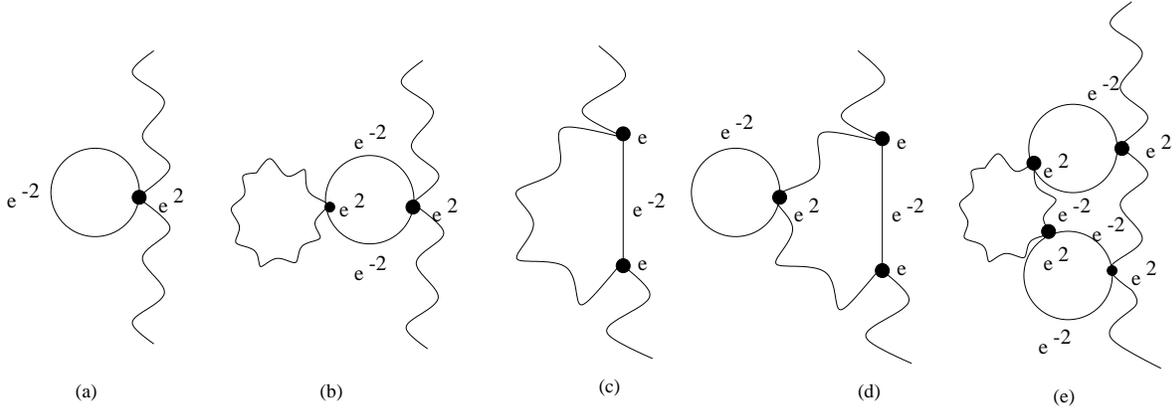}
\end{center}
\caption{Some 1PI self-energy diagrams appearing at 
lowest order $(e^{-2})^0$ in the regime $T\sim T^c_{E\leftrightarrow M}$.  
According to our quasi-particle approach diagrams (a), (b), (c), and (d) are allowed. Diagram 
(e) is forbidden since it can be obtained from a one-particle-reducible diagram by the additional 
exchanges of two TLM gauge bosons. 
TLM and TLH fields are denoted by wavy and straight lines, respectively,. }      
\end{figure}
For the following discussion of the strong-coupling regime 
we assume that magnetic monopole do not 
appear as propagating degrees of freedom. At $|e|\gg 1$ the TLH propagators 
reduce to $1/(m^{TLH}_E)^2\propto e^{-2}$. This leads to a power 
counting in $e^{-2}$ for each TLM self-energy diagram. Our discussion uses the DG. 
The nonrenormalizability of this gauge is not an issue in our effective theory. In DG 
loop integrals are cut off at momenta $\sim T$. The overall power of $e^{-2}$ belonging to a 
TLM self-energy diagram is never {\sl negative} which can easily be proven by induction. 
In Fig.\,3 some 1PI contributions to the radiatively induced TLM mass at lowest order in $e^{-2}$ are shown. 
It is not hard to see that the use of a one-loop resummed 
TLH propagator (summation of TLM-tadpoles) in diagrams $(a)$ and $(c)$ 
leads to radiatively generated contributions to the 
TLM mass $\sim \sqrt{N}T$. Corrections are controlled by powers 
of the small parameter $(N-1)^{-1} (T_E/T)^3$. It thus appears 
that the radiatively generated mass of TLM gauge bosons stays 
finite in the limit $|e|\to\infty$. To satisfy TSC with 
loop-induced masses for TLM modes is more involved than at tree-level. One 
can express loop-induced masses for TLM modes by 
$a^{TLH}_E$ and the gauge coupling $|e|$. As a consequence, the partial pressure 
exerted by TLH modes depends on $T$, $a^{TLH}_E$, and $|e|$. The latter can be expressed 
by $a^{TLH}_E$ using (\ref{Tdepe}). Imposing TSC (Eq.\,(\ref{sscE})) we see that 
radiative corrections lead to the occurrence of an additive correction to the right-hand 
side of (\ref{ladiffE}) due to the TLM modes becoming loop-induced QPEs. Although 
we have not investigated this term in detail we do not expect it to lead to a qualitative 
change of the tree-level behavior. After all the pre-factor of 
the associated integral is down by $\sim (N-1)^{-1}$ due 
to the smaller number of degrees of freedom 
associated with TLM modes. Clearly, 
an analysis involving higher 1PI irreducible loop orders 
is needed. We leave this analysis for future work.

\section{High temperature: Magnetic phase}

\subsection{Model and ground-state structure}

We have noticed in Sec.\,3.2 that the $N(N-1)$ TLH gauge bosons in 
the electric phase, which correspond to the broken generators of the 
coset SU($N$)/U(1)$^{N-1}$, decouple from the dynamics at the 
point $T_{E\leftrightarrow M}^c$. The surviving gauge symmetry 
is U(1)$^{N-1}$. This symmetry is 
spontaneously broken by the condensation of $(N-1)$ monopole species. 
We introduce a classical, complex scalar 
field $\phi_n$ for each of these condensates. There is a gauge 
field $B^{(n)}_\mu$ for the $n$th U(1) symmetry. This field is coupled 
with strength $g_n$ to the condensate $\phi_n$. 
The associated field strength $F^{(n)}_{\mu\nu}$ and 
the covariant derivative $D_\mu$ are defined as
\eqb
\label{covder}
F^{(n)}_{\mu\nu}=\pd_\mu B^{(n)}_\nu-\pd_\nu B^{(n)}_\mu\,,\ \ 
\ \ \ D^{(n)}_\mu=\pd_\mu+ig_n B^{(n)}_\mu\,.
\eqe
The Euclidean action at finite $T$ reads
\eab
\label{eucact}
S_M&=&-\int_0^{\beta} dx_4 \int d^3x\,\left\{\frac{1}{4}\sum_{n=1}^{N-1}
F^{(n)}_{\mu\nu}F^{(n)}_{\mu\nu}+\right.\nonumber\\ 
& &\left.
\frac{1}{2}\sum_{n=1}^{N-1}\left[\overline{D^{(n)}_\mu\phi_n}D^{(n)}_\mu\phi_n+V^{(n)}(|\phi_n|)\right]\right\}\,.
\eae
We define $V^{(n)}\equiv \bar{V}^{(n)}_{1/2}V^{(n)}_{1/2} $ with
\eqb
\label{pot}
\bar{V}^{(n)}_{1/2}=\e^{-i\delta_n}\frac{\La_M^3}{\bar{\phi}_n}\,,\ \ 
\ \ \ V^{(n)}_{1/2}=\e^{i\delta_n}\frac{\La_M^3}{\phi_n}\,
\eqe
where $\La_M$ denotes a mass scale. 
According to our general approach (see Sec.\,2) 
we ignore the gauge field sector when searching for zero-energy 
solutions of the scalar sector, which are periodic in $x_4$. 
The associated BPS equations read   
\eqb
\label{BPSeqM}
\pd_{4}\phi_n=\bar{V}^{(n)}_{1/2}\,,\ \ \ \ \ \ \pd_{x_4}\bar{\phi}_n=V^{(n)}_{1/2}\,.
\eqe
For $\delta_n\equiv\pi/2$ in (\ref{pot}) Eqs.\,(\ref{BPSeqM}) are solved by the following periodic functions
\eqb
\label{phin}
\phi_n(x_4)=\sqrt{\frac{\Lambda_M^3}{2\pi n T}}\exp\left[2\pi n i\,T x_4\right]\,.
\eqe
The choice of the winding number $n$ in (\ref{phin}) is in analogy to 
that in the electric case. For maximal symmetry breaking a minimal solution with unit increment of 
winding between neighboring SU(2) blocks was used in (\ref{persoln}), and (\ref{phin}) is 
just an implementation of this prescription in the magnetic phase.    

The nonfluctuating gauge fields of the ground state are zero-curvature 
solutions to the $(N-1)$ Maxwell equations in the backgrounds (\ref{phin}) 
\eqb
\label{Maxwell}
\pd_\mu F^{(n)}_{\mu\nu}=ig_n\left[\overline{D_\nu\phi_n}\phi_n-
\bar{\phi}_n D_\nu\phi_n\right]\,.
\eqe
We have 
\eqb
\label{Bsolution}
B^{(n)}_\nu=-\delta_{\nu 4}\frac{2\pi n}{g_n} T\,.
\eqe
The solutions (\ref{Bsolution}) imply the absence of 
the kinetic terms in (\ref{eucact}), and consequently 
the action density reduces to the potential term
\eqb
\label{BT}
\sum_{n=1}^{N-1} V^{(n)}(|\phi_n|)\equiv B_M(T)=\pi N(N-1)\La_M^3T\,.
\eqe

\subsection{Thermal self-consistency and quasi-particles}

The considerations concerning TSC are in close analogy to the tree-level approach to TSC in 
the electric phase. Again, non-Goldstone scalar modes 
are too heavy to be statistically relevant. We have
\eqb
\label{smass}
\frac{m_{s,n}}{T}\equiv\sqrt{\frac{1}{2}\,\pd^2_{|\phi_n|}V^{(n)}}/T=\sqrt{12}\pi\,n\,,\ 
\ \ \ (n=1,\cdots, N-1)\,.
\eqe
So the only fluctuating fields are 
the $(N-1)$ gauge fields. Their modes can be identified with 
noninteracting QPEs of temperature dependent masses $m_{M,n}=|g_n \phi_n|$. Due 
to the Abelian nature of the theory there are no radiative 
corrections to these masses induced by the quatum fluctuations of gauge bosons. 
For simplicity we assume that all gauge bosons 
masses are given by the lowest value, 
\eqb
\label{massimM}
m_{M,n}\equiv m_M\equiv |g_{N-1} \phi_{N-1}|\equiv|g| \sqrt{\frac{\La_M^3}{2\pi (N-1) T}}\,.
\eqe
This approximation leads to an upper bound for the pressure exerted 
by the QPEs in the magnetic phase. 
Similar to the electric phase the total pressure $p_M$ and the total energy density 
$\ep_M$ are given as
\eqb
\label{presseneM}
p_M=p_M^{id}-B_M(T)\,,\ \ \ \ \ \ep_M=\ep_M^{id}+B_M(T)\,
\eqe
We have defined
\eab
\label{pidM,eidM}
p^{id}_M&=&-\frac{3(N-1)T^4}{2\pi^2}P(a_M)\,,\ \ \ \ \ \ \ \ \ 
\ep^{id}_M=\frac{3(N-1)T^4}{2\pi^2}E(a_M)\,,\nonumber\\ 
a_M&\equiv&\frac{m_M}{T}\,.
\eae
The functions $B_M(T)$ and $P(a)$, $E(a)$ are defined in (\ref{BT}) 
and (\ref{pid,eid}), respectively.
TSC of the approach (\ref{presseneM}) 
is guaranteed if \cite{Gorenstein}
\eqb
\label{ctcM}
\frac{\pd p_M}{\pd m_M}=\frac{\pd p^{id}_M}{\pd m_M}-\frac{\pd B_M}{\pd m_M}=0\,.
\eqe
Eqs.\, (\ref{presseneM}), (\ref{pidM,eidM}), and (\ref{ctcM}) imply the 
following evolution equation 
\eqb
\label{ladiffM}
\frac{d\lambda_M}{da_M}=-\frac{12}{(2\pi)^6 N}\lambda_M^4\,a_M\,D(a_M)\,,
\eqe
where $\lambda_M\equiv T/T_M$, $T_M\equiv \La_M/(2\pi)$, 
and $a\,D(a)$ is defined in (\ref{defi}). 
Again, it has been assumed that $g$ depends on $T$ only {\sl implicitly} via 
$a_M$. An inverted solution to (\ref{ladiffM}), $a_M(\lambda_M)$, 
implies the following $\lambda_M$ dependence of the coupling $|g|$
\eqb 
\label{TdepeM}
|g|(\lambda_M)=\sqrt{\frac{(N-1)\lambda^3_M}{4\pi^2}}\,\,a_M(\lambda_M)\,.
\eqe
In analogy to the electric phase coupling constants $g_n$ diverge for 
$T\searrow T^c_{M\leftrightarrow C}\equiv \lambda^c_M T_M$. On a microscopic level, that is, 
for resolutions larger than $T$, the theory has Nielsen-Oleson 
vortices (NOVs) \cite{NO} for each 
of the $(N-1)$ spontaneously broken U(1)s. For small 
couplings $g_n\ll 1$ these vortices behave 
like classical objects. It is argued in \cite{NO} 
that in the limit of strong coupling $g_n\gg 1$ 
the action of a typical NOV
line (with a length comparable to the width) is $\propto g_n^{-2}$. 
This means that the energy carried by a typical NOV vanishes in this limit. 
At $g_n\gg 1$ NOVs are thus good candidates for CDOFs, and 
we expect that a transition to a phase occurs where they appear only 
collectively in condensed form.

\section{Low temperature: Broken center symmetry}

\subsection{Model and ground-state structure}

At $T=T^c_{M\leftrightarrow C}$ the remaining $(N-1)$ 
gauge bosons $B^{(n)}_\mu$ are infinitely heavy, and thus they decouple 
from the dynamics. Continuous 
gauge symmetry is reduced to a symmetry under the discrete 
subgroup of SU($N$): $Z_N$. The possibility has been discussed in 
\cite{KraussWilczek} that gauge theories masquerade as 
theories with a discrete symmetry when viewed from a low-energy 
observers perspective. We identify the center phase with the 
confined phase of the underlying SU($N$) gauge theory.

We have 
argued in Sec.\,4.2 that the termination of the magnetic phase is due to the 
condensation of NOVs. We introduce 
complex scalar fields $\Phi_n$, ($n=1,\cdots,N-1$),  
for the condensate of each vortex species and 
consider the following $Z_N$-symmetric action  
\eab
\label{eucactC}
S_C&=&-\int_0^{x_4} d\tau \int d^3x\,\frac{1}{2}\sum_{n=1}^{N-1}\left\{
\overline{\pd_\mu\Phi_n}\pd_\mu\Phi_n+v^{(n)}(\Phi_n)\right\}\,.
\eae
where $v^{(n)}\equiv\overline{v^{(n)}_{1/2}}v^{(n)}_{1/2}$. The functions $v^{(n)}_{1/2}$ and 
$\overline{v^{(n)}_{1/2}}$ are defined as
\eqb
\label{pot1/2}
v^{(n)}_{1/2}=\exp[i\delta_n]\left(\frac{\La_C^3}{\Phi_n}-\frac{\Phi_n^{N-1}}{\La_C^{N-3}}\right)\,,\ \ \ \ 
\overline{v^{(n)}_{1/2}}=\exp[-i\delta_n]\left(\frac{\La_C^3}{\bar{\Phi}_n}-
\frac{\bar{\Phi}_n^{N-1}}{\La_C^{N-3}}\right)\,,
\eqe
and $\La_C$ denotes the confinement scale. A nontrivial $Z_N$ transformation acts by a multiplication 
of each field $\Phi_n$ with one and the same unit root $\exp[\frac{2\pi il}{N}]$ ($l=1,\cdots, N-1$). Notice 
that only the pole part survives in (\ref{pot1/2}) for $|\Phi_n|<\La_C$ and $N\to\infty$. Periodic solutions to 
the BPS equations 
\eqb
\label{BPSeqC}
\pd_{x_4}\Phi_n=\overline{v^{(n)}}_{1/2}\,,\ \ \ \ \ \ \pd_{x_4}\bar{\Phi}_n=v^{(n)}_{1/2}\,.
\eqe
exist for $\delta_n=\pm\frac{\pi}{2}$. For $N\to\infty$ and $|\Phi_n|<\La_C$ 
they are given by the functions in (\ref{phin}). The action density on these solutions reads
\eqb
\label{BTC}
2\times\sum_{n=1}^{N-1} v^{(n)}(|\Phi_n|)\equiv B_C(T)=2\pi N(N-1)\La_C^3T=-p_C\,.
\eqe
For $N\to\infty$ the modulus of the fields $\Phi_n$ is locked at $|\Phi_n|=\La_C$ 
since the curvature of the potential at this point diverges, 
see (\ref{massotC}). Along the `trench' $|\Phi_n|=\La_C$ the potential 
is zero. Solutions $\Phi_n$ to (\ref{BPSeqC}) are 
no longer approximating the classical, Euclidean dynamics as in the cases with 
spontaneously broken {\sl continuous} gauge symmetry, they are exact. 
We observe that the masses of 
possible scalar QPEs are large for $N\to\infty$  
\eqb
\label{massotC}
m_{\Phi_n}\equiv\sqrt{\pd^2_{|\Phi_n|}{v^{(n)}}}=
\left\{\begin{array}{c}\sqrt{12}\pi\,n\,T\,,\ \ \ \ (|\Phi_n|<\La_C)\nonumber\\ 
\sqrt{2}\,N\,\La_C\,,\ \ \ \ (|\Phi_n|=\La_C)\end{array}\right.\,.
\eqe
These QPEs are thus thermodynamically irrelevant. Due to the exactness of the solutions $\Phi_n$ 
it should not matter whether we calculate the energy density $\ep_C$ 
in the classical field theory on the one hand or 
by applying the thermodynamic relation 
\eqb
\label{pressC}
\ep_C=T\frac{dp_C(T)}{dT}-p_C(T)
\eqe
to the pressure $p_C$ defined in (\ref{BTC}) on the other hand. Indeed, both approaches
yield $\ep_C=0$ for the nontrivial situation $|\Phi_n|<\La_C$: 
Field theoretically due to the BPS saturation of the $\Phi_n$, thermodynamically 
due to the linear dependence of $p_C$ on $T$. Notice that for $|\Phi_n|<\La_C$ 
the thermodynamic 
pressure is negative while the energy density vanishes. This 
situation is resolved by a violent 
relaxation of the fields $\Phi_n$ to $|\Phi_n|=\La_C$ where 
both pressure and energy density vanish. 

Things are more complicated for $N<\infty$. In this case there still are periodic 
solutions $\Phi_n$ to (\ref{BPSeqC}), see \cite{Hof}. However, the modulus $|\Phi_n|$ 
now depends on $x_4$. This implies an $x_4$-dependence of the masses 
$m_{\Phi_n}\equiv\sqrt{|\Phi_n|}{v^{(n)}}$. 
As a consequence a thermodynamic quasi-particle treatment fails. A real-time, 
non-equilibrium approach is needed to describe the relaxation of $\la \Phi_n\ra$ 
to one of the minima $\Phi^*_n=\La_C\,\exp[\frac{2\pi il}{N}]$. Notice that 
energy {\sl and} pressure vanish at $\Phi^*_n$. 
It is worth mentioning that the potential $\sum_{n=1}^{N-1} v^{(n)}$ 
in (\ref{eucactC}) has no inflexion point for $N<9$. In any case 
we expect a rapid relaxation on a time scale $\Delta t\sim \left.m^{-1}_{\Phi}\right|_{\Phi^*_n}=
\frac{\La_C}{\sqrt{2}\,N}$. This relaxation is accompanied 
by a violent (glue ball) particle production \cite{Brandenberger}.

\section{Matching the phases}

Let us now relate the scales $\La_E$, $\La_M$, 
and $\La_C$ by imposing the Clausius-Clapeyron condition that temperature and pressure are continuous 
across the phase boundaries.  

We first discuss the transition from the 
electric to the magnetic phase. At the critical 
temperature $T^c_{E\leftrightarrow M}$ the masses $m^{TLH}_{E,\alpha}$ of the $N(N-1)$ 
TLH gauge bosons diverge since the coupling constant $e$ diverges. Thus these degrees of freedom decouple 
from the dynamics. Their contribution to the total pressure vanishes. 
Since the couplings $g_n$ are {\sl inversely} proportional 
to the electric coupling $e$ and since the gauge boson masses in the magnetic phase are given as 
$m_{M,n}=|g_n\phi_n|$ we conclude that very close to $T^c_{E\leftrightarrow M}$ there are 
$(N-1)$ massless gauge bosons on either 
side of the transition point. This is an idealization since radiative
corrections do generate sizable masses for the TLM modes in the electric phase, see the 
discussion in Sec.\,3.3. For simplicity we assume tree-level 
behavior in the following \footnote{On the magnetic side gauge bosons start out 
massless. Thus there are 2 polarizations for $T\sim T^c_{M\leftrightarrow C}$. A 
smooth interpolation to a situation at lower $T$, where all gauge bosons 
are sufficiently massive to carry 
3 polarizations, must be operative.}. The condition 
\eqb
\label{E<->M}
p_E(T^c_{E\leftrightarrow M})=p_M(T^c_{E\leftrightarrow M})
\eqe
implies the following relation between $\La_E$ and $\La_M$
\eqb
\label{relEM}
\La_E=\La_M
\left\{\begin{array}{c}\left(16\,\frac{N-1}{N+2}\right)^{1/3} \,,\ \ \ \ (\mbox{even}\, N)\nonumber\\ 
\left(16\,\frac{N}{N+1}\right)^{1/3}\,,\ \ \ \ (\mbox{odd}\, N)\end{array}\right.\,.
\eqe
For the 
transition from the magnetic to the center 
symmetric phase at $T^c_{M\leftrightarrow C}$ no QPEs have to be considered 
if $N$ is sufficiently large. The reasons are diverging masses $m_{M,n}$ of the gauge 
bosons for $T\searrow T^c_{M\leftrightarrow C}$ and the strong Boltzmann suppression of scalar 
excitations in the center symmetric phase, see (\ref{massotC}). For large $N$ the condition         
\eqb
\label{M<->C}
p_M(T^c_{M\leftrightarrow C})=p_C(T^c_{M\leftrightarrow C})
\eqe
implies that
\eqb
\label{relMC}
\La_M=2^{1/3}\La_C\,.
\eqe
Eqs.\,(\ref{relEM}) and (\ref{relMC}) express the 
fact that there is 
only one freely adjustable mass scale 
describing the gauge theory cascade.   

\section{Numerical demonstration 
of the cascade}

In this section we numerically solve the evolution 
equations (\ref{ladiffE}) and  (\ref{ladiffM}) to 
demonstrate exemplarily the behavior which we relied upon 
in the discussions of 
the previous sections. In order to have 
some confidence in (\ref{relMC}), 
which only at large $N$ is a good approximation, 
we set $N=10$ in the following. For small values of $N$, as they occur in the 
SU($N$) factors of the 
standard model, in technicolor models, and GUTs, the matching 
between the magnetic and the 
center phases can, in principle, 
be performed numerically.    

We start with a calculation of the running QPE mass 
in the electric phase. For $N=10$ the evolution equation (\ref{ladiffE}) takes 
the following form
\eqb
\label{ladiffE10}
\frac{d\lambda_E}{da^{TLH}_E}=-\frac{9}{(2\pi)^6}\lambda_E^4\,a^{TLH}_E\,D(a^{TLH}_E)\,.
\eqe
We assume that $a^{TLH}_E=0$ at the initial 
temperature $T_E^i=\frac{\lambda_E(a^{TLH}_E=0)}{2\pi}\La_E$. 
Fig.\,4 shows a numerical solution to (\ref{ladiffE10}) 
for $\lambda_E(0)=100$. 
\begin{figure}
\begin{center}
\leavevmode
\leavevmode
\vspace{8cm}
\includegraphics{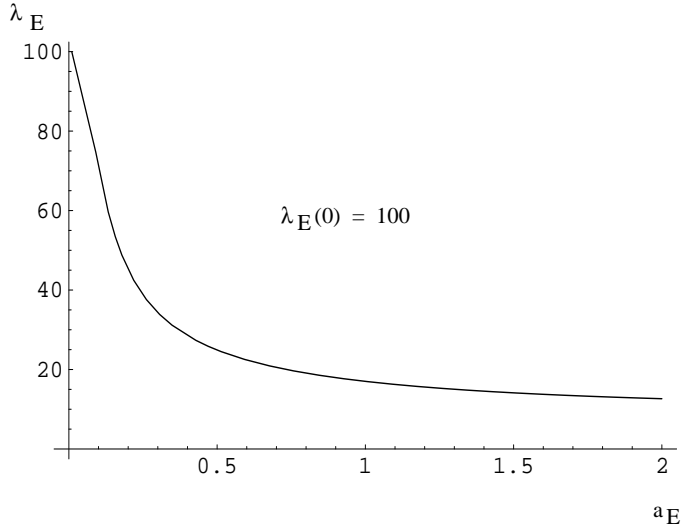}
\end{center}
\caption{The numerical solution to (\ref{ladiffE10}) for $\lambda_E(0)=100$.}   
\end{figure}
Keeping in mind the tree-level relation (\ref{Tdepe}), we notice that the coupling $e$ 
grows strong ($|e|>1$) well before saturation. This saturation occurs at $\lambda^c_E\sim 10$, 
see Fig.\,4. The (weak) dependence on the initial conditions of the critical
temperature $T^c_{E\leftrightarrow M}=\frac{\lambda^c_E}{2\pi}\La_E$ is demonstrated in Fig.\,5.
 
Let us now discuss the magnetic phase. For $N=10$ the 
evolution equation (\ref{ladiffM}) takes the following form
\eqb
\label{ladiffM10}
\frac{d\lambda_M}{da_M}=-\frac{6}{5}\frac{1}{(2\pi)^6}\lambda_M^4\,a_M\,D(a_M)\,.
\eqe
According to (\ref{relEM}) the initial value $\lambda_M(0)$ is related to 
the critical value $\lambda^c_E$ as
\eqb
\label{inicrit}
\lambda_M(0)=\lambda^c_E \left\{\begin{array}{c}\left(16\frac{N-1}{N+2}\right)^{1/3} \,,
\ \ \ \ (\mbox{even}\, N)\nonumber\\ 
\left(16\frac{N}{N+1}\right)^{1/3}\,,\ 
\ \ \ (\mbox{odd}\, N)\end{array}\right.
\eqe
\begin{figure}
\begin{center}
\leavevmode
\leavevmode
\vspace{8cm}
\includegraphics{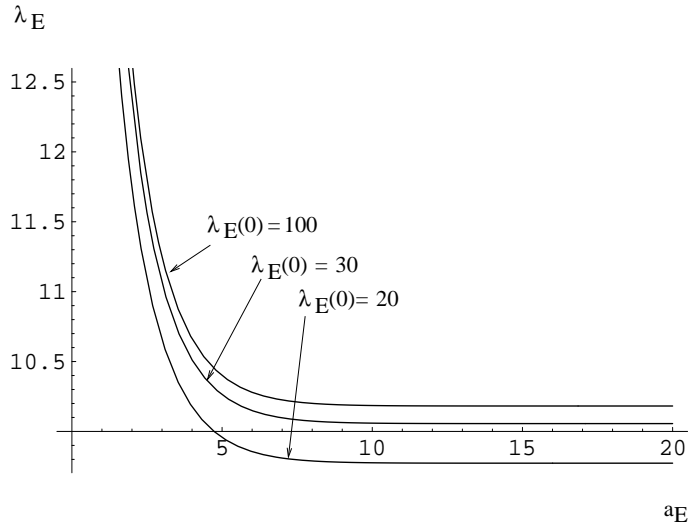}
\end{center}
\caption{Dependence of the saturation value $\lambda_E^c$ on the initial conditions.}   
\end{figure}
For $N=10$ this leads to
\eqb
\label{conEM}
\lambda_M(0)=(12)^{1/3}\lambda^c_E\sim 2.3\,\lambda^c_E\,.
\eqe
Fig.\,5 suggests a value of $\lambda^c_E=10$. Using (\ref{conEM}) 
we may take $\lambda_M(0)=23$ as an initial value 
for the evolution of $\lambda_M$. The corresponding solution to 
(\ref{ladiffM10}) is shown in Fig.\,5. Notice the smallness of the 
temperature range belonging to the magnetic phase: saturation 
of $\lambda_M$ (or the transition to the center phase) occurs at about $T^c_{M\leftrightarrow C}\sim
\frac{17}{23}\, T^c_{E\leftrightarrow M}\sim 0.74\, T^c_{E\leftrightarrow M}$! Fig.\,6 suggests the following 
value for $\lambda_M^c$
\eqb
\label{crit}
\lambda_M^c=17=2\pi \frac{T^c_{M\leftrightarrow C}}{\La_M}\,.
\eqe
Finally, by virtue of (\ref{relMC}) and (\ref{crit}) the critical temperature 
$T^c_{M\leftrightarrow C}$ can be expressed in terms of the confinement scale $\La_C$ as
\eqb
\label{confs}
T^c_{M\leftrightarrow C}\sim 3.4\, \La_C\,,\ \ \ \ (N=10)\,.
\eqe
At the temperatures $T^n_C=\frac{\La_C}{2\pi n}$ the field moduli $|\Phi_n|$ would get 
locked in the limit $N\to\infty$. The lowest value $T^1_C=\frac{\La_C}{2\pi}$ is 
much smaller than $T^c_{M\leftrightarrow C}$. According to (\ref{phin}) 
we have $|\Phi_n|\ll \La_C$ for $T\gg T^1$. 
Thus the potential is dominated by its pole parts, and 
the use of the matching relation (\ref{relMC}) is consistent.
\begin{figure}
\begin{center}
\leavevmode
\leavevmode
\vspace{8cm}
\includegraphics{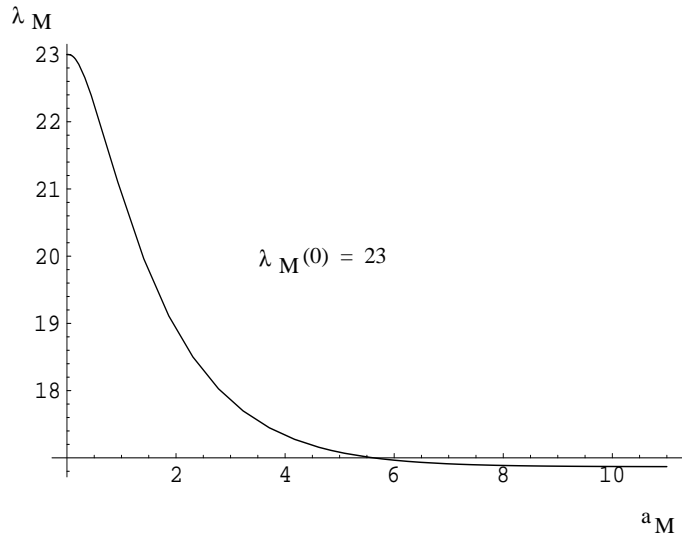}
\end{center}
\caption{The solution to (\ref{ladiffM10}) for $\lambda_M(0)=23$. This value of $\lambda_M(0)$ matches 
the evolution in the electric phase.}   
\end{figure}

\section{Summary and outlook}

The results of SU($N$) lattice gauge theory simulations 
of the thermodynamic potentials indicate a considerable deviation 
from the ideal-gas behavior even for 
temperatures larger than $T_c^{dec}$ \cite{lattice}. These effects are not captured 
by naive, unsummed thermal perturbation theory which by poor convergence
predicts its own breakdown for experimentally feasible temperatures \cite{Iancu}. The purpose of this paper 
was to construct an alternative approach to hot SU($N$) gauge dynamics. 
It is based on the assumption that away from a phase 
transition the bulk of the quantum interaction in the system is residing in the 
collective properties of condensed low-energy degrees of freedom. Incomplete classical dynamics 
was assumed to describe this condensed part. This assumption turned 
out to be self-consistent. To satisfy in addition 
the constraints imposed by thermodynamic equilibrium, 
namely zero gauge curvature and coordinate independent 
pressure of the ground state, $\phi^{-2}$-type 
potentials for the Higgs-field VEVs $\{\phi\}$ were required. 

Due to the spontaneous gauge symmetry breaking, which is 
induced by the condensed part, (some of) the statistically 
fluctuating degrees of freedom become quasi-particle excitations at tree-level. 
The very existence of quasi-particles was 
necessary for the thermodynamic self-consistency of our approach. 
The nonperturbative evolution of the gauge coupling(s) with temperature 
is a consequence of thermodynamic self-consistency. This evolution is from weak to strong coupling 
as temperature decreases. Generically, the situation at very large coupling(s) is as follows: 
On the one hand, topological defects, which behave quasi-classically at small coupling(s), become 
quasi-massless, and thus they condense. Tree-level massless modes pick up sizable masses due to 
radiative corrections. On the other hand, tree-level quasi-particles
acquire a large masses, they 
decouple from the dynamics. As a result, a transition 
to a phase with reduced (gauge) symmetry occurs. This process may repeat itself in an 
underlying theory with a large non-Abelian gauge symmetry. In a final transition at 
low temperature the system reaches a 
phase with a (spontaneously broken) discrete symmetry. 

Specifically, for the SU($N$) pure gauge theory considered in this paper 
the system starts out in an 'electric' phase. In this phase the 
ground state is macroscopically described by a classical adjoint Higgs 
field which may consist of condensed calorons. It was observed in Sec.\,3.1 
that the associated ground-state solution is in agreement with asymptotic freedom; 
the Higgs field `melts' with increasing temperature. In the electric phase the evolution 
with decreasing temperature runs into a transition 
to a 'magnetic' phase. This transition reduces the gauge symmetry 
from SU($N$) to U(1)$^{N-1}$ at some critical temperature $T^c_{E\leftrightarrow M}$. 
In the new phase the ground state is made of condensed magnetic monopoles.  
In a final transition at $T^c_{M\leftrightarrow C}<T^c_{E\leftrightarrow M}$ the system 
condenses magnetic vortices. This reduces the continuous gauge symmetry 
U(1)$^{N-1}$ to the discrete symmetry Z$_N$. This last phase is the confining one, 
see also \cite{centerlat} for lattice 
investigations of the SU(2) case. 

The present paper demonstrates that the description 
of hot SU($N$) gauge dynamics in terms of an (incomplete) 
classical ground-state solution and noninteracting quasi-particles can be useful. 
The deviation from the Stefan-Boltzmann 
limit of the thermodynamic pressure measured on the 
lattice at large temperature is qualitatively explained. A negative equation 
of state is inevitable for temperatures close to $T^c_{M\leftrightarrow C}$ 
which can explain vacuum energy on cosmological scales 
both in the very early universe and today. In the former case an explanation of 
the seeding of large-scale structure during inflation requires an additional 
light field -- the curvaton \cite{LythWands}. This field is generated dynamically if the 
underlying gauge theory has fundamentally charged, chiral fermions \cite{Hof2002}. 
Looking at the accelerated expansion of 
todays universe in the light of the present paper, one would conclude  
that the near coincidence of the vacuum and dark matter energy density 
is due to a gauge theory that is close to a transition 
to its discrete-symmetry phase.    

Much remains to be done to adapt strategy of the present paper to realistic 
low-energy theories such as 
QCD. In the following we list three points which seem to be most urgent.
\begin{itemize}
\item 
Finite $N$ center-phase transition: 
The transition to the center phase and the subsequent 
evolution towards one of the minima of the potential 
at finite $N$ is not accessible to a thermodynamic treatment. 
Real-time calculations along the lines of refs.\,\cite{Brandenberger} 
are needed to compute the rates of particle number creation in the early stages of this transition 
and the late-time behavior of the confining phase of the 
system. 
\item  
Radiatively induced masses for TLM modes: 
In the end of the electric phase the gauge-coupling modulus 
$|e|$ is large and quantum effects are explicit. 
The TLH gauge bosons acquire large masses and thus cease to be 
statistically relevant. Their quantum fluctuations induce sizable 
masses for the TLM gauge bosons which are 
expanded in orders of $e^{-2}$. An all-loop orders result for 
the lowest-order, $(e^{-2})^{0}$, would be desirable.
\item 
Inclusion of fundamentally charged fermions: It is not possible to consistently 
define a local quantum theory of propagating electric and 
magnetic charges. In our approach, however, magnetic monopoles are either decoupled 
due to large masses (weakly coupled regime of electric phase), 
or they appear in condensed form without 
collective excitations (magnetic phase). 
The situation may be problematic only 
for $T\searrow T^c_{E\leftrightarrow M}$ where for a short period magnetic 
monopoles may be explicit degrees of freedom. Thus for a large range of temperatures 
we can not see a principle objection against the inclusion of 
fundamental, chiral fermions into our approach. There are, however, a number of 
open questions. For example, gauge invariance 
puts no strong constraint on the way how quarks couple to the 
respective condensates. The simplest coupling would be of the Yukawa-type. 
The thermodynamic self-consistency of the approach then would lead to a 
coupled set of evolution equations. Would chiral symmetry breaking and the 
{\sl dynamical decoupling} of {\sl isolated} quarks simply be the statement 
that these Yukawa couplings diverge during the transition to the center phase?

\end{itemize}

\section*{Acknowledgments}
The author would like to thank J. Berges, H. Gies, and Z. Tavartkiladze 
for useful conversations.

\bibliographystyle{prsty}

\end{document}